\documentclass[aps,reprint,showpacs,floatfix]{revtex4-1}
\usepackage{amsmath,graphicx,color,verbatim}

\begin{document}

\title{Prediction of Li intercalation voltages in rechargeable battery cathode materials: effects of exchange-correlation functional, van der Waals interactions, and Hubbard $U$}
\author{Eric B. Isaacs}
\author{Shane Patel}
\author{Chris Wolverton}
\email{c-wolverton@northwestern.edu}
\affiliation{Department of Materials Science and Engineering, Northwestern University, Evanston, Illinois 60208, USA}

\begin{abstract}
  Quantitative predictions of the Li intercalation voltage and of the
  electronic properties of rechargeable battery cathode materials are
  a substantial challenge for first-principles theory due to the
  possibility of (1) strong correlations associated with localized
  transition metal $d$ electrons and (2) significant van der Waals
  (vdW) interactions in layered systems, both of which are not
  accurately captured by standard approximations to density functional
  theory (DFT). Here, we perform a systematic benchmark of electronic
  structure methods based on the widely-used generalized gradient
  approximation of Perdew, Burke, and Ernzerhof (PBE) and the new
  strongly constrained and appropriately normed (SCAN)
  meta-generalized gradient approximation for battery cathode
  materials. Studying layered Li$_x$TiS$_2$, Li$_x$NiO$_2$, and
  Li$_x$CoO$_2$, olivine Li$_x$FePO$_4$, and spinel Li$_x$Mn$_2$O$_4$,
  we compute the voltage, crystal structure, and electronic structure
  with and without extensions to incorporate on-site Hubbard
  interactions and vdW interactions. Within pure DFT (i.e., without
  corrections for on-site Hubbard interactions), SCAN is a significant
  improvement over PBE for describing cathode materials, decreasing
  the mean absolute voltage error by more than 50\%. Although explicit
  vdW interactions are not critical and in cases even detrimental when
  applied in conjunction with SCAN, Hubbard $U$ corrections are still
  in general necessary to achieve reasonable agreement with
  experiment. We show that no single method considered here can
  accurately describe the voltage and overall structural, electronic,
  and magnetic properties (i.e., errors no more than 5\% for voltage,
  volume, band gap, and magnetic moments) of battery cathode
  materials, motivating a strong need for improved electronic
  structure approaches for such systems.
\end{abstract}

\date{\today}
\maketitle

\section{Introduction}\label{sec:intro}


Li-ion rechargeable battery cathodes, which are typically composed of
transition metal oxides, represent a challenging testbed for
first-principles theory. Prediction of the Li intercalation voltage is
of particular importance given it is a fundamental battery property
that helps determine the battery power and closely relates to the
(de)lithiation mechanism. The average Li intercalation voltage $V$
relates to the difference in Li chemical potential between the cathode
and anode. For a battery cathode whose Li content changes changes from
$x_1$ to $x_2>x_1$, $V$ relative to Li metal is given by
$eV=E_{\mathrm{Li}} + \frac{E(x_1)-E(x_2)}{(x_2-x_1)},$ where $e$ is
the elementary charge, $E_{\mathrm{Li}}$ is the energy of Li metal,
and $E(x)$ is the energy of the cathode material with Li concentration
$x$ \cite{aydinol_ab_1997,wolverton_cation_1998}. For example, for
Li$_x$CoO$_2$, over the full range of Li ($x_1=0$ for CoO$_2$, $x_2=1$
for LiCoO$_2$), $V$ is given by $eV = E_{\mathrm{Li}}+E(0)-E(1)$. In
principle, Gibbs free energies should be used in the previous
expressions; we ignore pressure-volume and entropic contributions
typically small compared to the magnitude of $V$
\cite{reynier_entropy_2004}. Since $V$ is a function of the total
energies of phases whose electron distributions differ starkly
(leading to fewer opportunities for error cancellation), it is a
useful observable to serve as a stringent benchmark of \textit{ab
  initio} thermodynamics approaches.

Density functional theory (DFT)
\cite{hohenberg_inhomogeneous_1964,kohn_self-consistent_1965}, within
the generalized gradient approximation (GGA)
\cite{perdew_generalized_1996} in particular, has become the
\textit{de facto} standard for electronic structure calculations of
solids. Despite its many successes, DFT struggles to capture the
composition-dependent energetics necessary to describe the
intercalation voltage and compositional phase stability of battery
cathode materials. For example, in the case of olivine Li$_x$FePO$_4$
\cite{padhi_phospho-olivines_1997,padhi_effect_1997}, DFT in the GGA
substantially underestimates the experimental $V$ (by $\sim$20\%) and
fails to qualitatively capture the experimentally-observed phase
separation for intermediate Li concentrations
\cite{zhou_phase_2004,zhou_first-principles_2004}. In order to address
this challenge, a wide variety of electronic structure approaches,
including DFT with different exchange-correlation functionals
\cite{aydinol_ab_1997,wolverton_cation_1998,wolverton_first-principles_1998,wang_computational_2016,chakraborty_predicting_2018},
hybrid functionals \cite{chevrier_hybrid_2010,seo_calibrating_2015},
van der Waals (vdW) functionals \cite{aykol_van_2015}, DFT plus
on-site Hubbard $U$ (DFT+$U$)
\cite{zhou_first-principles_2004,bacq_impact_2004,ong_voltage_2011,aykol_local_2014,shishkin_self-consistent_2016,isaacs_compositional_2017},
DFT+$U$+$V$ (where $V$ is an inter-site interaction)
\cite{cococcioni_energetics_2019}, DFT plus dynamical mean-field
theory \cite{isaacs_compositional_2019}, and diffusion quantum Monte
Carlo \cite{saritas_charge_2020} have been applied to battery cathode
voltages. It remains an open question what interactions and level of
theory beyond DFT in the GGA are needed to adequately describe such
materials.


Among these works, we mention two recent developments pertaining to
cathode voltage prediction. The first is the work of Aykol \textit{et
  al.}, who found that vdW interactions, employed in conjunction with
DFT+$U$ and the widely-used GGA of Perdew, Burke, and Ernzerhof (PBE),
are necessary to accurately describe the voltage for layered
Li$_x$CoO$_2$ \cite{aykol_van_2015}. Second, also studying layered
cathode materials, Chakraborty \textit{et al.} found that the new
strongly constrained and appropriately normed (SCAN) DFT functional
considerably improves the voltage prediction even without Hubbard $U$
or explicit vdW corrections \cite{chakraborty_predicting_2018}.

In this work, we perform a systematic study of the average Li
intercalation voltage of five classic cathode materials: layered
Li$_x$TiS$_2$, olivine Li$_x$FePO$_4$, layered Li$_x$NiO$_2$, spinel
Li$_x$Mn$_2$O$_4$, and layered Li$_x$CoO$_2$. For the five cathode
materials, we investigate the impact on $V$ of (1) the new SCAN
exchange-correlation functional, (2) the use of density functionals
explicitly considering vdW interactions, and (3) on-site Hubbard $U$
within DFT+$U$. In all cases, the calculations are critically compared
to experiments to assess accuracy. We also consider other quantities
(volumes, electronic band gaps, and magnetic moments) in order to
provide a more complete picture of the accuracy of the methods.

Within pure DFT \footnote{We use ``pure DFT'' to refer to all the
  methodologies studied in this work without Hubbard $U$ corrections,
  such that the total energy is purely a functional of the density
  (even if only implicitly). This includes SCAN and the vdW
  functionals.}, we find that SCAN is a significant improvement for
describing battery cathode materials, decreasing the
mean-absolute-error for $V$ by more than 50\%, as compared to PBE
\cite{perdew_generalized_1996}, from 0.67 V to 0.30 V.
In some cases (e.g., Li$_x$FePO$_4$ and Li$_x$Mn$_2$O$_4$) Hubbard $U$
corrections are still necessary to achieve reasonable agreement (i.e.,
within 5\%) with experiment. Therefore, given the improvement of SCAN
over PBE, quantitative voltage predictions within pure DFT are closer
but still currently out of reach. When applied in combination with
SCAN, Hubbard $U$ within DFT+$U$ can lead to worsened predictions in
some cases (i.e., Li$_x$TiS$_2$ and Li$_x$CoO$_2$). In other words,
within DFT+$U$, Hubbard $U$ added to SCAN does not universally help or
hurt the predictions. In the one case in which SCAN itself provides a
sufficient prediction of $V$ (i.e., Li$_x$NiO$_2$), we find it is
still possible to achieve a better overall electronic structure
description using DFT+$U$ calculations based on PBE rather than SCAN.
Therefore, calculations based on SCAN should not necessarily be
considered universally better than those based on PBE when considering
both the energetics and overall electronic structure. Using PBE,
adding vdW interactions provide appreciably improved $V$ predictions,
even for non-layered cathode materials lacking a clear van der Waals
gap when delithiated. In the majority of the cases, when Hubbard $U$
is also considered with PBE, the experimental $V$ can be achieved with
or without vdW interactions (for different values of $U$). Therefore,
it is not clear that missing vdW interactions are a significant source
of PBE's well-known voltage underprediction in general. We find that
such vdW interactions are significantly less important in terms of $V$
when applied in conjunction with SCAN, which already contains some
intermediate-range vdW interactions. Overall, despite the
significantly improved $V$ predictions of SCAN as compared to PBE
within pure DFT, we illustrate that no single method considered here
can generally describe the voltage and overall electronic structure of
battery cathode materials, motivating a strong need for improved
electronic structure approaches for such systems.

\section{Summary of electronic structure approaches tested}\label{sec:methodology}

\subsection{Exchange-correlation functional}

The key ingredient to DFT is the exchange-correlation functional
$E_{xc}$, which encapsulates all the interaction effects beyond the
single-particle kinetic energy and mean-field (Hartree) Coulomb energy
\cite{martin_electronic_2008}. $E_{xc}$ in the local density
approximation (LDA) depends solely on the electron density $\rho$ and
is parametrized to exactly describe the homogeneous electron gas
(jellium) \cite{ceperley_ground_1980}. In GGA, $E_{xc}$ depends on
$\nabla\rho$ in addition to $\rho$, which allows for the satisfaction
of additional constraints such as the correct behavior in the slowly-
and rapidly-varying density limits \cite{perdew_generalized_1996}. A
higher level of theory is the meta-GGA, in which $E_{xc}$ exhibits an
additional dependence on the orbital kinetic energy
density \begin{equation}\tau = \sum_i \frac{1}{2}
  |\nabla\psi_i|^2,\end{equation} where $\psi_i$ is the $i$th occupied
Kohn-Sham wavefunction, corresponding to functional that is implicitly
nonlocal in $\rho$.

The strongly constrained and appropriately normed (SCAN) functional
\cite{sun_strongly_2015} is a new meta-GGA, which satisfies 17 known
constraints of the exact $E_{xc}$ and has shown significant promise in
the description of solids
\cite{sun_strongly_2015,sun_accurate_2016,tran_rungs_2016,kitchaev_energetics_2016,zhang_comparative_2017}.
Just as PBE is built on top of LDA (reproducing the LDA result for
jellium), SCAN is built on top of PBE and exhibits the same behavior
as PBE for slowly-varying densities in the metallic bonding regime of
$\tau$. We note that the SCAN functional implicitly contains some
``intermediate-range'' vdW interactions \cite{sun_strongly_2015}, and
it also can be incorporated in methods containing explicit vdW
interactions \cite{peng_versatile_2016}, as discussed below.

Very relevant to the prediction of battery cathode voltages is that
SCAN has been shown to yield significant improvement to formation
energy predictions as compared to PBE for strongly-bound compounds
like oxides
\cite{isaacs_performance_2018,zhang_efficient_2018}. Indeed, for a few
layered cathode materials, very recent work has suggested that SCAN
achieves more accurate $V$ prediction compared to PBE
\cite{chakraborty_predicting_2018,isaacs_compositional_2019}. In
particular, based on calculations of layered Li$_x$NiO$_2$,
Li$_x$CoO$_2$, and Li$_x$MnO$_2$ with PBE, PBE+$U$, and SCAN,
Chakraborty \textit{et al.} found that SCAN performs better than PBE
and PBE+$U$ for the $V$ profiles. Based on the $V$ behavior, as well
as predicted lattice parameters, densities of states, and $\rho$ (as
compared to that from the PBE0 hybrid functional), they concluded that
SCAN without Hubbard $U$ exhibits good overall performance for layered
cathode materials. Whether such trends hold more generally (e.g., for
non-layered cathodes) is an open question addressed by this work.

\subsection{Explicit van der Waals interactions}

The lack of nonlocal correlation effects needed to capture vdW
interactions is a well-documented limitation of standard DFT
\cite{grimme_density_2011,klimes_perspective_2012,berland_van_2015,grimme_dispersion-corrected_2016,hermann_first-principles_2017,stohr_theory_2019}.
In order to address this limitation, first-principles vdW density
functionals have been developed. In such functionals, a nonlocal
correlation energy term (explicitly nonlocal in $\rho$) of the form
\begin{equation}E_c^{nl}=\frac{1}{2}\int\int \rho(r) \phi(r, r')
  \rho(r') d^3rd^3r'\end{equation} is incorporated in $E_{xc}$
\cite{dion_van_2004,roman-perez_efficient_2009}. Here, $\phi(r, r')$
is the kernel, which is typically based on approximations to the
frequency-dependent polarizability. For example, Dion \textit{et al.}
devised a kernel based on a plasmon pole approximation to the
dielectric function $\epsilon$ and a second-order expansion of the
polarization $S=1-\epsilon^{-1}$
\cite{dion_van_2004,grimme_dispersion-corrected_2016}, such that the
kernel is a function of $\rho$ and $\nabla\rho$ at spatial coordinates
$r$ and $r'$ as well as $|r-r'|$.

Aykol \textit{et al.} recently tested a variety of methodologies
incorporating vdW interactions (including first-principles and
semiempirical approaches) on Li$_x$CoO$_2$ \cite{aykol_van_2015}. The
first-principles opt-type vdW density functionals
\cite{klimes_chemical_2010,klimes_van_2011}, such as optPBE-vdW, were
found to yield the most accurate $V$ predictions and correspond to a
significant improvement over standard density functionals lacking vdW
interactions. This opens up the question of how important vdW
interactions are to describe battery cathode materials in general
(i.e., beyond Li$_x$CoO$_2$), which we address in this work.

We focus on optPBE-vdW in this work
\cite{klimes_chemical_2010,klimes_van_2011}. optPBE-vdW combines a
linear combination of the exchange forms of PBE and the related RPBE
\cite{hammer_improved_1999}, LDA local correlation, and the kernel of
Ref. \citenum{dion_van_2004}. In optPBE-vdW, the fraction of PBE-like
\cite{perdew_generalized_1996} and RPBE-like
\cite{hammer_improved_1999} exchange and the two parameters employed
in both such forms have been optimized (hence the ``opt'') to minimize
interaction energy errors for the Set 22 (S22) quantum chemistry
benchmark \cite{jurecka_benchmark_2006}. When we refer to adding vdW
interactions to PBE in this work, we are referring to the optPBE-vdW
method. Although this is not strictly accurate as the difference
between PBE and optPBE-vdW is not additive, we do so for convenience
and since optPBE-vdW is closely connected to PBE. We also consider the
SCAN plus revised Vydrov-Van Voorhis 2010 (SCAN+rVV10) vdW functional
\cite{vydrov_nonlocal_2010,sabatini_nonlocal_2013,peng_versatile_2016},
which corresponds to a different choice of kernel with one of its two
parameters fit to best reproduce the Ar dimer binding curve from
coupled cluster singles, doubles, and perturbative triples [CCSD(T)]
quantum chemistry calculations. SCAN+rVV10 is explored in this work
for purely practical reasons as it is currently the only vdW
functional implemented in conjunction with the SCAN functional in the
Vienna \textit{ab initio} simulation package (\textsc{vasp}). In this
work, we also refer to SCAN+rVV10 as SCAN+vdW for convenience.

\subsection{On-site Hubbard $U$ corrections}

In an attempt to correct for the deficiencies of DFT (using common
approximations like the GGA), the DFT+$U$ approach
\cite{himmetoglu_hubbard-corrected_2014} has become a widely used
method to describe cathode materials. In this methodology, DFT is
augmented with an on-site Hubbard interaction $U$ (solved within
static mean-field theory) related to strong electronic correlations in
a chosen subspace of localized orbitals (defined via transition metal
$d$ orbital projectors in this work). In this methodology, the energy
depends on the on-site density matrix for the transition metal $d$
orbitals in addition to $\rho$. In particular, using the simplified
rotationally-invariant formalism of Dudarev \textit{et al.}
\cite{dudarev_electron-energy-loss_1998} and the fully localized limit
(FLL) double counting \cite{anisimov_density-functional_1993}, the
DFT+$U$ energy can be written as
\begin{equation}E_{DFT+U} = E_{DFT}[\rho^s] + \frac{1}{2}U\sum_{\tau,
    m, s}n_m^{\tau s}(1-n_m^{\tau s}),\label{eq:dftu}\end{equation}
where $\rho^s$ is the spin-density, $E_{DFT}[\rho^s]$ is the
(spin-dependent) DFT energy and $n_m^{\tau s}$ is the $m$th eigenvalue
of the density matrix corresponding to transition metal site $\tau$
and spin projection $s$.  Written this way, it is visible that the
effect of DFT+$U$ is to penalize non-integer occupancy of the
localized orbitals.

DFT+$U$ has been shown to help alleviate the voltage underestimation
of DFT in the GGA \cite{zhou_first-principles_2004}, and it has become
a standard tool to describe cathode materials. However, recent
evidence suggests it may lead to considerable problems. In particular,
for Li$_x$CoO$_2$, DFT+$U$ yields spurious gaps and charge ordering,
as well as overestimated Li order-disorder temperatures
\cite{isaacs_compositional_2017}. The ability of DFT+$U$ to accurately
describe cathode materials in general is an open question we aim to
address in this work. We note that DFT+$U$ is a static approximation
to the more accurate DFT plus dynamical mean-field theory (DFT+DMFT),
in which the local correlation problem is solved exactly rather than
via the Hartree-Fock approximation of DFT+$U$. Recent DFT+DMFT
calculations found a significantly different $V$ prediction for
Li$_x$CoO$_2$ as compared to DFT+$U$, suggesting dynamical
correlations neglected by DFT+$U$ but captured by DFT+DMFT may also be
important in battery cathode materials
\cite{isaacs_compositional_2019}. However, due to the large
computational cost to solve the quantum impurity problem in DFT+DMFT,
we do not explore the role of dynamical correlations in this work.

\section{Computational Details}\label{sec:compdetails}

Spin-dependent density functional theory calculations using the
projector augmented wave (PAW) method
\cite{blochl_projector_1994,kresse_ultrasoft_1999} and a 520 eV plane
wave kinetic energy cutoff are performed using \textsc{vasp}
\cite{kresse_ab_1994,kresse_ab_1993,kresse_efficient_1996,kresse_efficiency_1996}.
We use the Perdew--Burke--Ernzerhof (PBE) GGA
\cite{perdew_generalized_1996} and the strongly constrained and
appropriately normed (SCAN) \cite{sun_strongly_2015} meta-GGA to the
exchange-correlation functional. The impact of vdW interactions is
assessed via calculations with the optPBE-vdW functional
\cite{klimes_chemical_2010,klimes_van_2011} and the SCAN+rVV10
functional \cite{peng_versatile_2016}. On-site Hubbard $U$ is included
for the transition metal $d$ states using the rotationally-invariant
DFT+$U$ approach
\cite{liechtenstein_density-functional_1995,himmetoglu_hubbard-corrected_2014}.
We use the recommended \textsc{vasp} 5.2 PBE PAW potentials for all
calculations \cite{vasp_paw}. Uniform $k$-meshes are chosen with
$\ge 8,000/N_{\mathrm{atoms}}$ $k$-points, where $N_{\mathrm{atoms}}$
is the number of atoms in the unit cell. The ionic forces and total
energy are converged to 10$^{-2}$ eV/\AA\ and 10$^{-6}$ eV,
respectively. We employ 0.1 eV 1st-order Methfessel-Paxton smearing
\cite{methfessel_high-precision_1989} for structural relaxations and
the tetrahedron method with Bl\"{o}chl corrections
\cite{blochl_improved_1994} for static runs.

\begin{figure*}
  \includegraphics[width=\linewidth]{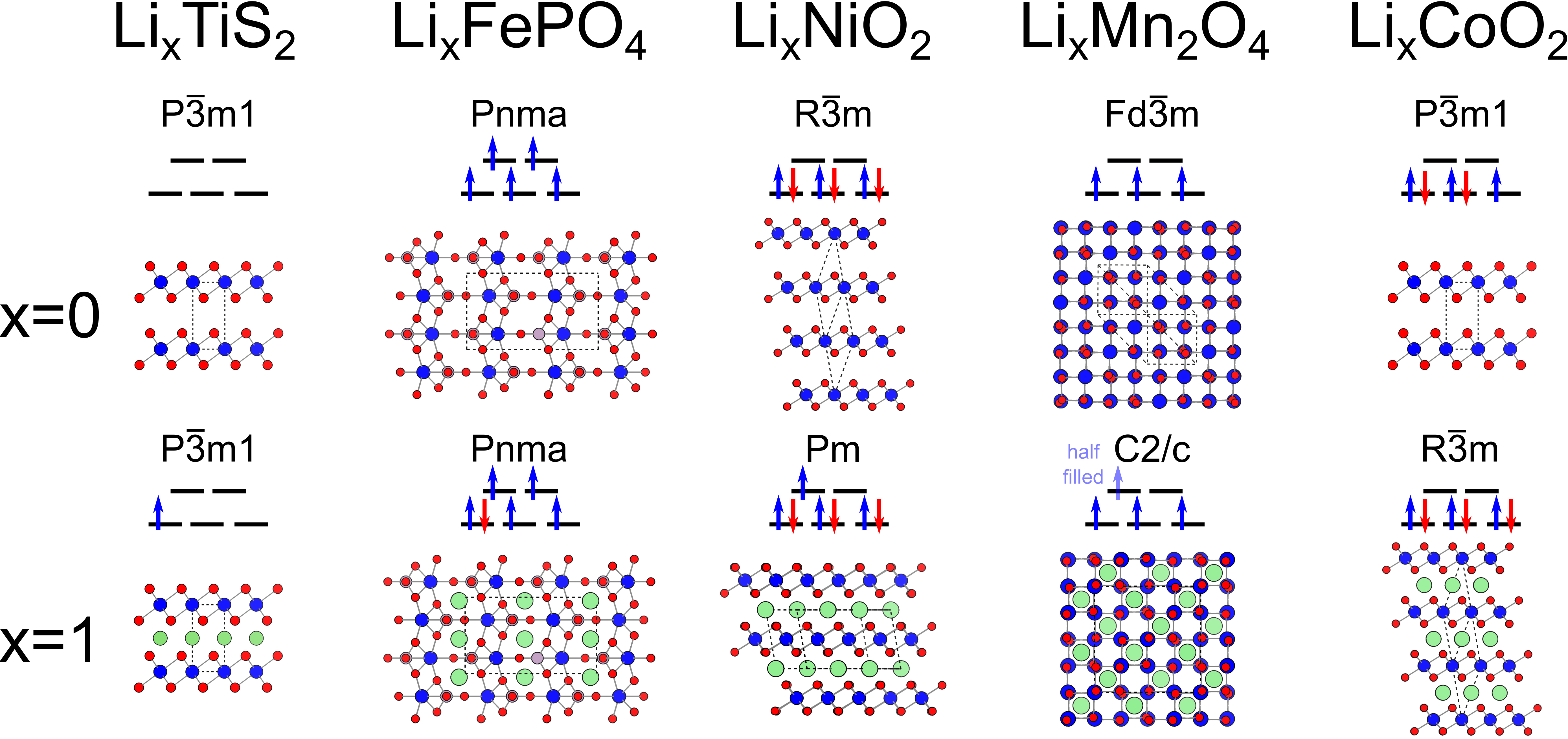}
  \caption{Crystal structures for the fully delithiated (top, $x=0$)
    and fully lithiated (bottom, $x=1$) cathode materials studied in
    this work. Green, blue, red, and purple circles correspond to Li,
    transition metal, oxygen, and phosphorus atoms, respectively, and
    the unit cell is indicated by black dashed lines. Above each
    crystal structure, the space group and nominal transition metal
    $d$ electronic configuration, assuming $1+$ Li oxidation state and
    $2-$ oxidation state, are given. For simplicity, we only show the
    octahedral crystal field splitting into the 3 lower-energy
    $t_{2g}$ levels and 2 higher-energy $e_g$ levels. In the case of
    LiMn$_2$O$_4$, the $e_g$ manifold is nominally occupied by 1/2 an
    electron per Mn on average, as indicated by the ``half filled''
    label.\label{fig:crystal_structures}}
\end{figure*}

\section{Results and Discussion}\label{sec:results}

\subsection{Crystal structures and nominal electronic configurations}

We begin by briefly discussing the structures and electronic
configuration of the cathode materials considered. The crystal
structures and nominal transition metal (TM) $d$ electron
configurations for the five cathode materials are illustrated in Fig.
\ref{fig:crystal_structures}. All the compounds considered have
octahedral coordination of TM by oxygen. Although the octahedra are
often distorted, we still refer to the lowest energy three $d$ levels
as $t_{2g}$ and the highest energy two $d$ levels as $e_g$ for
simplicity, where $t_{2g}$ and $e_g$ are the irreducible
representations of $d$ orbitals in perfect octahedral symmetry.

Li$_x$TiS$_2$, Li$_x$NiO$_2$, and Li$_x$CoO$_2$ are layered materials
with alternating layers of Li and edge-sharing TM--oxygen octahedra.
TiS$_2$ ($t_{2g}^0e_{g}^0$), LiTiS$_2$ ($t_{2g}^1e_{g}^0$), and
CoO$_2$ ($t_{2g}^5e_{g}^0$) are in the hexagonal $P\bar{3}m1$
structure (O1 structure)
\cite{chianelli_structure_1975,dahn_structure_1980,amatucci_coo2_1996}.
NiO$_2$ and LiCoO$_2$ (both $t_{2g}^6e_{g}^0$) are considered in the
rhombohedral $R\bar{3}m$ structure (O3 structure)
\cite{johnston_preparation_1958,orman_cobaltiii_1984,seguin_structural_1999,croguennec_nio2_2000}.
We model LiNiO$_2$ ($t_{2g}^6e_{g}^1$) with the monoclinic $Pm$
structure \cite{cao_local_2009}, which captures the Jahn-Teller
distortion. For the Li$_x$CoO$_2$ case, in addition to computing $V$
for the full $0 < x < 1$ range, we also compute $V$ for
$x<\frac{1}{2}$ and $x>\frac{1}{2}$. To do so, we consider
Li$_{1/2}$CoO$_2$ in the known monoclinic $P2/m$ structure, which
corresponds to an in-plane Li/vacancy ordering in a unit cell twice as
large as the primitive rhombohedral cell
\cite{takahashi_single-crystal_2007}.

In contrast to the other materials, Li$_x$FePO$_4$ and
Li$_x$Mn$_2$O$_4$ do not exhibit layered crystal structures. Olivine
FePO$_4$ ($t_{2g}^3e_{g}^2$) and LiFePO$_4$ ($t_{2g}^4e_{g}^2$)
crystallize in an orthorhombic $Pnma$ structure consisting of (1)
one-dimensional channels of Li and (2) layers of corner sharing
Fe--oxygen octahedra connected via phosphate groups
\cite{santoro_antiferromagnetism_1967,rousse_magnetic_2003}. To model
Li$_x$Mn$_2$O$_4$, we consider Mn$_2$O$_4$ ($t_{2g}^3e_{g}^0$) in the
ideal spinel-like $Fd\bar{3}m$ structure ($\lambda$-MnO$_2$), which
consists of a diamond sublattice of Li and a three-dimensional network
of edge-sharing Mn--oxygen octahedra
\cite{hunter_preparation_1981,mosbah_phases_1983}. In order to capture
possible Jahn-Teller effects, we model LiMn$_2$O$_4$ (nominally in the
$t_{2g}^{3.5}e_{g}^0$ configuration) with the symmetry-broken
ferromagnetic monoclinic $C2/c$ structure from Ref.
\citenum{kim_surface_2015}. Ferromagnetic ordering is considered for
all magnetic compounds except Li$_x$FePO$_4$, which exhibits
antiferromagnetic ordering \cite{rousse_magnetic_2003}.

\subsection{Pure DFT}

\begin{figure}[htpb]
  \includegraphics[width=\linewidth]{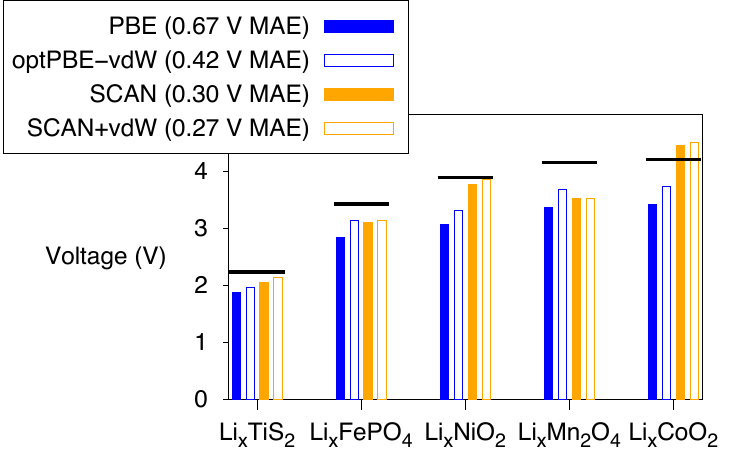}
  \caption{Average intercalation voltage for $0 < x < 1$ within
    PBE, optPBE-vdW, SCAN, and SCAN+vdW for DFT (i.e., $U=0$). The
    solid black horizontal lines indicate the experimental
    voltage. Mean absolute error (MAE) values for the 5 cathode
    materials are indicated in the legend.\label{fig:dft_voltages}}
\end{figure}

Figure \ref{fig:dft_voltages} shows the average intercalation voltages
over the full Li concentration range ($0 < x < 1$) within pure DFT
($U=0$). As has been shown previously
\cite{zhou_first-principles_2004}, PBE systematically and
substantially underpredicts $V$, yielding a mean absolute error (MAE)
of 0.67 V. SCAN represents a significant improvement over PBE in terms
of the predicted $V$, reducing the MAE by over 50\% to 0.30 V.
However, some errors are still unacceptably large (e.g., 15\% error
for Li$_x$Mn$_2$O$_4$). In this sense, quantitative $V$ predictions
within DFT are closer to being achieved but are still currently out of
reach. For most of the cathode materials, SCAN still underpredicts the
experimental values despite the appreciable increase in $V$ with
respect to PBE values. There are two exceptions to this trend: (1)
Li$_x$NiO$_2$, for which the SCAN prediction (3.8 V) is nearly
identical to the experimental value (3.9 V) and (2) Li$_x$CoO$_2$, for
which the SCAN prediction (4.5 V) is appreciably larger than
experiment (4.2 V). The increase in $V$ of SCAN with respect to PBE is
highly system dependent: while this increase is 1.0 V for
Li$_x$CoO$_2$, it is a mere 0.2 V for Li$_x$Mn$_2$O$_4$.

\begin{figure}[htbp]
  \begin{center}
    \includegraphics[width=1.0\linewidth]{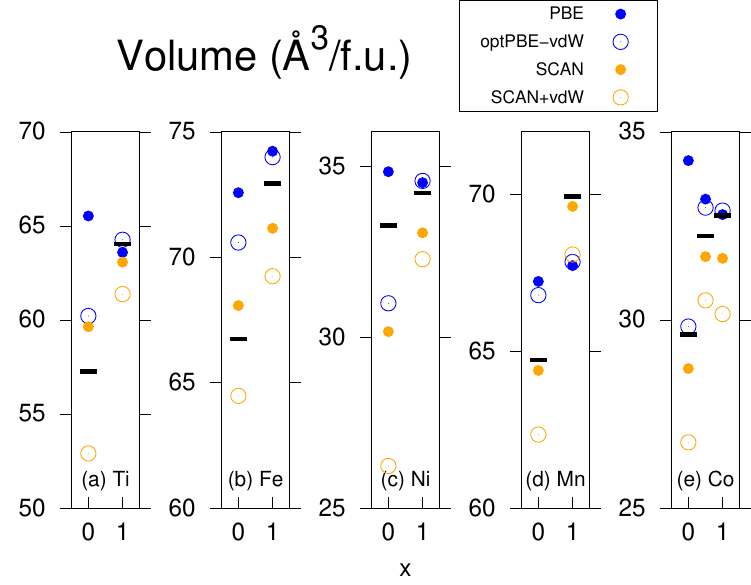}
  \end{center}
  \caption{Volumes in \AA$^3$ per formula unit within pure DFT for (a)
    Li$_x$TiS$_2$, (b) Li$_x$FePO$_4$, (c) Li$_x$NiO$_2$, (d)
    Li$_x$Mn$_2$O$_4$, and (e) Li$_x$CoO$_2$ as a function of Li
    concentration for the various methods considered in this work. The
    panels are labeled by the transition metal. Experimental values
    are shown as black horizontal lines
    \cite{whittingham_lithium_1975,mizushima_lixcoo2_1980,hunter_preparation_1981,hirano_relationship_1995,amatucci_coo2_1996,rousse_magnetic_2003,takahashi_single-crystal_2007,cao_local_2009}.}
  \label{fig:dft_volume}
\end{figure}

Explicit vdW interactions also generally yield an increase in
predicted $V$, though of a smaller magnitude. For example, adding vdW
interactions to PBE (i.e., optPBE-vdW) reduces the MAE from 0.67 V to
0.42 V. Here, the voltage increases are less system-dependent: similar
increases of 0.1--0.3 V (4--10\%) for optPBE-vdW with respect to PBE
are found for all five cathode materials. The $V$ enhancement is not
generally smaller for non-layered materials: $V$ increases by 10\% for
Li$_x$FePO$_4$, for example. In contrast, adding vdW interactions to
SCAN (i.e., SCAN+vdW) does not appreciably increase predicted $V$ and
an MAE of 0.27 V (negligibly smaller than the 0.30 V value for SCAN)
is obtained. We believe this behavior stems from the construction of
SCAN+vdW since a parameter in the rVV10 form is fit specifically for
SCAN, which already intrinsically contains some intermediate-range vdW
interactions. Based on the predicted $V$ behavior, we find that
explicit vdW methods are not critical when applied in conjunction with
SCAN for battery cathode materials. This suggests that the vdW
interactions intrinsic to SCAN are likely sufficient to describe vdW
interactions in this class of materials. We note that Chakraborty
\textit{et al.} reached a similar conclusion using a distinct
dispersion-corrected DFT approach \cite{chakraborty_predicting_2018}.

The predicted volume is another observable with which we can benchmark
different computational methods. As shown in Fig.
\ref{fig:dft_volume}, we find SCAN+vdW leads to worsened volume
predictions compared to SCAN for all the systems considered. This
suggests that the explicit vdW interactions contained within SCAN+vdW
may be not only unnecessary, but even harmful to the description of
battery cathode materials. This behavior is in contrast to that of
optPBE-vdW, which generally improves the volume predictions as
compared to PBE.

We note that the impact of the explicit vdW interactions on $V$ is not
primarily structural in nature. For example, freezing to PBE ground
state structures, the SCAN $V$ value for Li$_x$CoO$_2$ changes by only
24 meV relative to the value calculated using the SCAN ground state
structures. We find similar behavior for the other cathode materials.
For example, for Li$_x$TiS$_2$, the computed $V$ changes by at most
0.1 V for the case with structures relaxed with vdW interactions and
that with structures relaxed without vdW interactions, for all the
functionals considered.

\begin{figure}[htpb]
  \includegraphics[width=\linewidth]{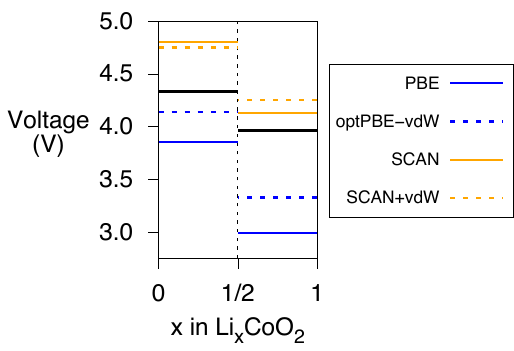}
  \caption{Average Li$_x$CoO$_2$ intercalation voltage for $0 < x <
    1/2$ (left horizontal lines) and $1/2 < x < 1$ (right horizontal
    lines) within PBE, optPBE-vdW, SCAN, and SCAN+vdW for DFT (i.e.,
    $U=0$). The solid black horizontal lines indicate the
    corresponding experimental values.\label{fig:dft_half_voltages}}
\end{figure}

For Li$_x$CoO$_2$, we also consider the separate ``half voltages,''
i.e., the distinct voltage averages for $0 < x < 1/2$ and $1/2 < x <
1$, shown in Fig. \ref{fig:dft_half_voltages} for DFT.  In
Fig. \ref{fig:dft_half_voltages}, one can observe the same main trends
discussed above for the Li$_x$CoO$_2$ $V$ over the full range of $x$:
(1) SCAN significantly increases the voltage, exceeding experiment and
(2) incorporating explicit vdW interactions moderately enhances the
voltages when added to PBE, but negligibly when added to SCAN. We
focus on the ``voltage gap'' at $x=1/2$, i.e., the difference between
the voltage average of $1/2 < x < 1$ and that of $0 < x < 1/2$. Such a
voltage gap $\Delta V$ is a measure of the formation energy of a
stable (on the convex hull) phase of intermediate Li concentration
with respect to the $x=0$ and $x=1$ endmembers, which can be written
as $-x(1-x)e\Delta V$ \cite{aykol_local_2014}. Therefore, the $x=1/2$
voltage gap of Li$_x$CoO$_2$ is a convenient benchmark for
compositional phase stability. The voltage gap predicted by PBE
($-0.9$ V) is significantly larger in magnitude than the experimental
value ($-0.4$ V). SCAN predicts an improved, but still too large (in
magnitude) voltage gap of $-0.7$ V. This is consistent with the
conclusion that SCAN provides an improved, though still imperfect,
description of the energetics of battery cathode materials. vdW
interactions also improve the predicted $\Delta V$, yielding values of
$-0.8$ V for optPBE-vdW and $-0.5$ V for SCAN+vdW.

\subsection{DFT+$U$}

\begin{figure*}
  \includegraphics[width=\linewidth]{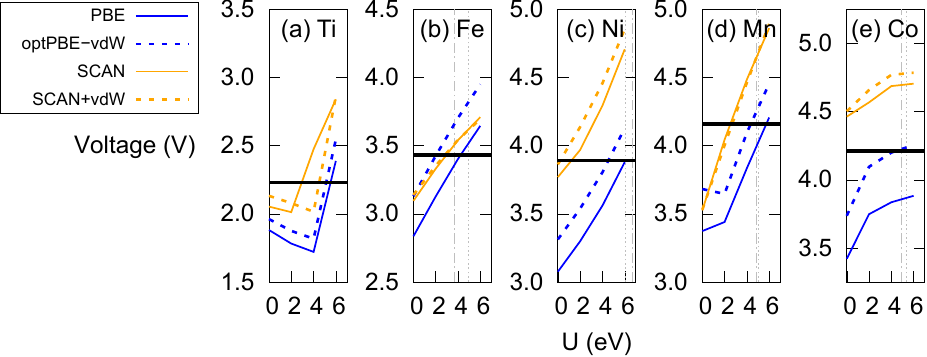}
  \caption{Average intercalation voltage as a function of $U$ for $0 <
    x < 1$ within PBE, optPBE-vdW, SCAN, and SCAN+vdW for DFT+$U$ for
    (a) Li$_x$TiS$_2$, (b) Li$_x$FePO$_4$, (c) Li$_x$NiO$_2$, (d)
    Li$_x$Mn$_2$O$_4$, and (e) Li$_x$CoO$_2$. The solid black
    horizontal lines indicate the experimental voltage.
    First-principles $U$ values from PBE linear response calculations
    \cite{zhou_first-principles_2004,cococcioni_linear_2005} are
    included when available as grey dotted ($x=0$) and dot-dashed
    ($x=1$) lines.\label{fig:dftu_voltages}}
\end{figure*}

Figure \ref{fig:dftu_voltages} shows the DFT+$U$ average intercalation
voltages over the full Li concentration range. We first comment on the
general impact of $U$ on the intercalation voltages. Although an
increase in $V$ has typically been found with increasing $U$ for
battery cathode materials (shown here as well as in previous works
\cite{zhou_first-principles_2004,zhou_phase_2004,aykol_local_2014,aykol_van_2015,isaacs_compositional_2017,cococcioni_energetics_2019}),
we also find the opposite behavior in the small-$U$ limit in some of
the cases (e.g., Li$_x$TiS$_2$). An increase in $V$ with $U$
necessarily stems from the larger energy penalty on the $x=0$
endmember than the $x=1$ endmember, since $eV$ is proportional to
$E(0) - E(1).$ Similarly, a decrease in $V$ with $U$ corresponds to a
greater energetic penalty on $x=1$ than $x=0.$

Nominal electron counting corresponding to completely filled or
completely empty states (as suggested by most of the level diagrams in
Fig. \ref{fig:crystal_structures}) is insufficient by itself to
explain these trends, as the energy penalty from DFT+$U$ (using the
FLL double counting) exactly vanishes in such a fully localized limit,
as can be seen in Eq. \ref{eq:dftu}. However, knowledge of the
electron counting in conjunction with the overall electronic structure
can be used to explain the observed trends. For example, for
Li$_x$CoO$_2$, $U$ penalizes metallic $x=0$ more than the band
insulator $x=1$, which has closer-to-integer $d$ orbital occupations
\cite{isaacs_compositional_2017} due to electron counting (as well as
to the increased ionicity stemming from Li).  Therefore, $V$ increases
with $U$. The reverse situation occurs for Li$_x$TiS$_2$: here, $x=0$
is the band insulator and $x=1$ has a partially-filled $t_{2g}$ shell,
corresponding to a metal. This explains the decrease in $V$ in the
small-$U$ limit (the increase at larger $U$ is discussed
later). Analogously, a negative $\partial V/\partial U$ in the
small-$U$ limit is also found for Li$_x$Mn$_2$O$_4$ using PBE and
optPBE-vdW since within these levels of theory Mn$_2$O$_4$ is
insulating and LiMn$_2$O$_4$ is metallic for small $U$. The voltage
increases with $U$ for Li$_x$FePO$_4$ and Li$_x$NiO$_2$, though the
origin of the increases is different than the Li$_x$CoO$_2$ case. For
Li$_x$FePO$_4$, whose endmembers are both magnetic insulators, it is
the enhanced covalency of the $x=0$ endmember
\cite{isaacs_compositional_2017} that gives it a larger energy
penalty. And for Li$_x$NiO$_2$, despite the nominally partially-filled
$e_g$ shell for $x=1$, the increased ordering from the Jahn-Teller
distortion allows the $x=1$ endmember to be less affected by $U$ than
the $x=0$ endmember.

\begin{figure*}[tb]
  \begin{center}
    \includegraphics[width=1.0\linewidth]{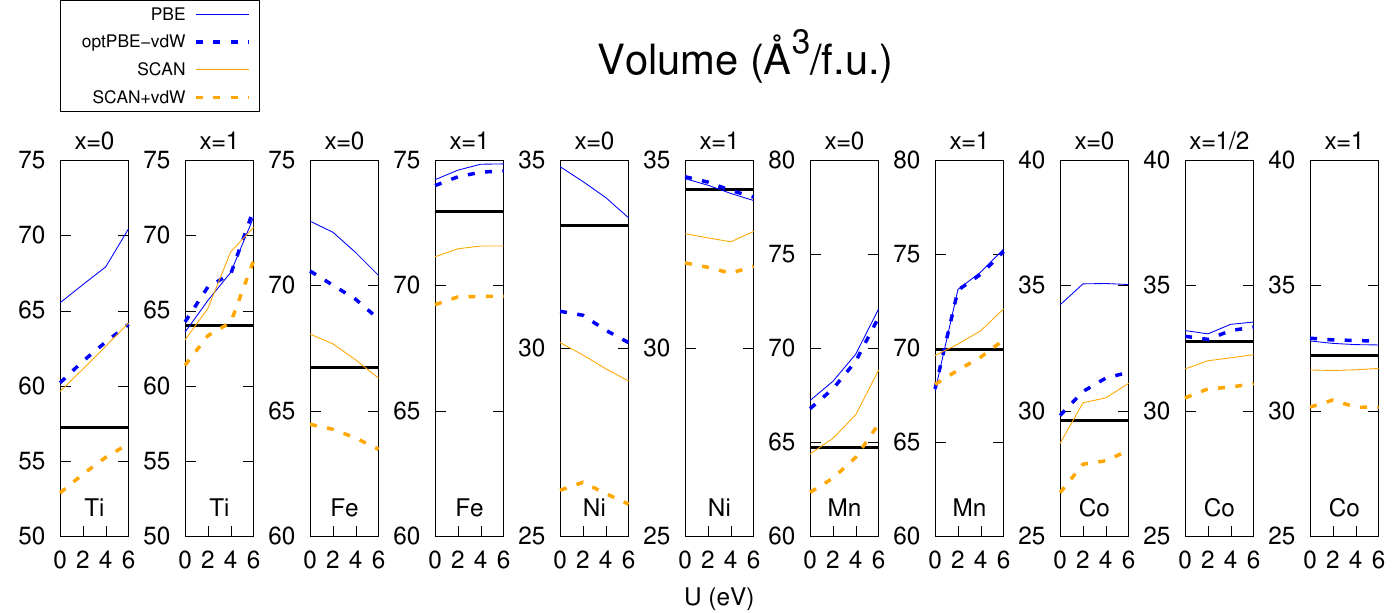}
  \end{center}
  \caption{Volumes in \AA$^3$ per formula unit within DFT+$U$ for
    Li$_x$TiS$_2$, Li$_x$FePO$_4$, Li$_x$NiO$_2$, Li$_x$Mn$_2$O$_4$,
    and Li$_x$CoO$_2$ as a function of $U$ for the various methods
    considered in this work. The panels are labeled by the transition
    metal. and Li concentration. Experimental values are shown as
    black horizontal lines
    \cite{whittingham_lithium_1975,mizushima_lixcoo2_1980,hunter_preparation_1981,hirano_relationship_1995,amatucci_coo2_1996,rousse_magnetic_2003,takahashi_single-crystal_2007,cao_local_2009}.}
  \label{fig:dftu_volume}
\end{figure*}

Due to the diversity of behavior observed, we now discuss the specific
results (including the behavior with different levels of theory in
comparison with experiment) for each material separately. Afterwards,
we present a general synthesis of the results.

We begin with Li$_x$TiS$_2$. As discussed above, small values of $U$
for Li$_x$TiS$_2$ serve to decrease the predicted $V$, which already
underestimates the experimental value within pure DFT. We also note
that, for the majority of the pure DFT levels of theory considered
here, $U$ also hurts the volume prediction (shown in Fig.
\ref{fig:dftu_volume}) for Li$_x$TiS$_2$. Therefore, in the case of
Li$_x$TiS$_2$, adding Hubbard $U$ serves to hurt the description.
Consistent with this finding, we note that the past work of Chevrier
\textit{et al.} avoided the use of Hubbard $U$ for Ti-based compounds
\cite{chevrier_hybrid_2010}.

Here, we comment on the discontinuous behavior at larger $U$. A
metal-insulator transition for LiTiS$_2$ at larger $U$ decreases its
energy penalty relative to TiS$_2$, leading to a change in sign in
$\partial V/\partial U.$ The insulating behavior is spurious as
LiTiS$_2$ is actually metallic in experiment
\cite{klipstein_transport_1987}. This raises the question of whether
this predicted insulating state for LiTiS$_2$ for larger $U$
corresponds to (1) an intrinsic failure of DFT+$U$ or simply (2) the
use of an unphysically-large $U$ parameter. Using a self-consistent
linear response approach and PBE, Shishkin and Sato computed $U$ of
5.5 eV for LiTiS$_2$ (using same the PAW potential for Ti we employ,
which treats the $4s$ semicore states as valence states)
\cite{shishkin_self-consistent_2016}. Since Shishkin and Sato found
LiTiS$_2$ to still be metallic at this $U$ value
\cite{shishkin_self-consistent_2016}, their results suggest the second
case above (too-large $U$); however, we note that this may be a
borderline case as we find insulating LiTiS$_2$ for $U=6$ eV.

In conjunction with the pure DFT results discussed above, we find that
the application of SCAN (as opposed to PBE) and vdW interactions
significantly improves the voltage prediction for Li$_x$TiS$_2$, while
$U$ hurts the description. We note that SCAN+vdW, which exhibits the
best agreement with the experimental $V$ (error of only 0.1 V), has
worse volume predictions than those of SCAN for Li$_x$TiS$_2$ (for
LiTiS$_2$ in particular), as shown in Fig. \ref{fig:dft_volume}.

For Li$_x$FePO$_4$, the pure DFT approaches are insufficient to
quantitatively describe the voltage. However, the predicted voltage
increases roughly linearly with $U$ in all cases, enabling agreement
with experiment using DFT+$U$. For PBE, the optimal $U$ value to
achieve agreement with experiment is 4.2 V, in agreement with previous
work
\cite{zhou_phase_2004,zhou_first-principles_2004,isaacs_compositional_2017}.
This value also agrees well with the overall magnitude of the
first-principles $U$ values for the $x=0$ (4.9 V) and $x=1$ (3.7 V)
endmembers computed from first principles (with PBE) via the linear
response approach \cite{zhou_first-principles_2004}. Since adding vdW
interactions to PBE (i.e., optPBE-vdW) provides a roughly rigid
increase in the predicted $V$, of around 0.3 V, the optimal $U$ value
to achieve agreement with the experimental $V$ for optPBE-vdW is 2.0
eV, substantially lower than the PBE case. SCAN+$U$ and SCAN+vdW+$U$
yield essentially identical $V$ predictions, consistent with the
intrinsic vdW interactions in SCAN. For such methods, the predicted
$V$ matches experiment for $U=3.0$ eV, also significantly lower than
the PBE case.

Although not computed here, it would be interesting to assess whether
the first-principles $U$ values based on optPBE-vdW and/or SCAN(+vdW)
would also be appreciably lower than those of PBE, leading to the same
consistency observed for PBE in terms of the first-principles $U$ and
$U$ fit to experimental $V$. We note that the optimal $U=3.0$ eV for
SCAN(+vdW), in terms of $V$, agrees well with $U$ values found to
reproduce the FeO/Fe$_2$O$_3$ (2.9 eV) and FeO/Fe$_3$O$_4$ (3.3 eV)
experimental oxidation reaction energies in a recent SCAN+$U$ work by
Sai Gautam and Carter \cite{sai_gautam_evaluating_2018}.

The volume behavior for Li$_x$FePO$_4$ is shown in Fig.
\ref{fig:dftu_volume}. SCAN+$U$ yields the best volume prediction for
Li$_x$FePO$_4$ among all the methods considered in this work, though
some underestimation of the LiFePO$_4$ volume persists. The band gap
and local Fe magnetic moment behaviors for Li$_x$FePO$_4$ are shown in
the Supplemental Material \footnote{See Supplemental Material for
  additional calculation results (magnetic moment, band gap, and
  density of states) and details on the experimental voltage data.}.
The application of $U$ to SCAN also significantly improves the
predicted LiFePO$_4$ band gap, though the FePO$_4$ gap (already in
good agreement with experiment for $U=0$) becomes overestimated. A
similar effect is found in terms of the local Fe magnetic moment, with
overestimation (underestimation) for FePO$_4$ (LiFePO$_4$).
Ultimately, while $U$ can be chosen to yield agreement with the
experimental $V$ using DFT+$U$ based on any of the pure DFT
methodologies considered here, we find that SCAN+$U$ using $U$ of
$\sim 3$ eV provides the best (although still imperfect) overall
description of Li$_x$FePO$_4$ when also taking into account the
volume, band gap, and local magnetic moments.


For Li$_x$NiO$_2$, the pure DFT prediction using PBE significantly
underestimates experiment, but agreement can be reached for $U$ of
$\sim 6$ eV, which is close in value to the PBE first-principles
computed endmember $U$ values \cite{zhou_first-principles_2004}. The
behavior is similar for optPBE-vdW+$U$, whose $V$ predictions are
$\sim 0.2$ eV larger than those of PBE+$U$. The behavior for SCAN is
quite distinct. Here, the SCAN-predicted $V$ already exhibits
excellent agreement (within $\sim 0.1$ V) with experiment even without
Hubbard $U$. Therefore, the application of $U$ in this case pushes $V$
to far too large values. This is also true for SCAN+vdW+$U$, which
exhibits a small, roughly constant $\sim 0.1$ V increase in $V$ with
respect to SCAN+$U$.

Although one can achieve a satisfactory quantitative $V$ prediction
using PBE/optPBE-vdW with $U$ ($\sim 6$ eV) or SCAN(+vdW) without $U$,
the volume prediction (shown in Fig. \ref{fig:dftu_volume}) suggests
such approaches are not equivalent in their overall description. SCAN
and SCAN+vdW provide worsened volume predictions as compared to
PBE+$U$. Although optPBE-vdW+$U$ yields a similar LiNiO$_2$ volume as
PBE+$U$, its volume prediction for NiO$_2$ is significantly worse than
PBE+$U$. We note additionally that PBE+$U$ yields an accurate band gap
prediction for LiNiO$_2$, as shown in the Supplemental Material.
Overall, despite the excellent $V$ prediction using SCAN(+vdW), we
find PBE+$U$ provides the overall best description of Li$_x$NiO$_2$.


The Li$_x$Mn$_2$O$_4$ case is similar to that of Li$_x$FePO$_4$ in
that a quantitatively accurate $V$ prediction can be achieved using
calculations based on any of the pure DFT methodologies considered
here, but only using Hubbard $U$. $U$ values of 5.7 eV (reasonably
close in value to the first-principles computed endmember $U$ values
\cite{zhou_first-principles_2004}) and 4.4 eV are needed to achieve
agreement with the experimental $V$ for PBE and optPBE-vdW,
respectively. The SCAN+$U$ and SCAN+vdW+$U$ voltages agree with
experiment for the significantly smaller value of $U=2.6$ eV. This
value is in good agreement with the $U$ values found to reproduce the
MnO/Mn$_2$O$_3$ (2.9 eV) and Mn$_2$O$_3$/MnO$_2$ (2.5 eV) experimental
oxidation reaction energies in the work of Sai Gautam and Carter
\cite{sai_gautam_evaluating_2018}.

Here, as in the Li$_x$FePO$_4$ case, the SCAN+vdW $V$ result is nearly
identical to that of SCAN. This suggests that the energetic impact of
the vdW interactions in SCAN+vdW beyond those already contained within
SCAN itself is especially small for the non-layered cathode materials.
In contrast, optPBE-vdW+$U$ yields a substantially larger $V$
prediction than PBE+$U$. Based on the volume data shown in Fig.
\ref{fig:dftu_volume} and band gap data shown in the Supplemental
Material, we find that DFT+$U$ calculations based on SCAN exhibit a
better overall description than those based on PBE for
Li$_x$Mn$_2$O$_4$. In particular, SCAN+$U$ and SCAN+vdW+$U$ do not
exhibit the significant volume overestimation of PBE+$U$ and
optPBE-vdW+$U$ for appreciable $U$. In addition, the LiMn$_2$O$_4$
band gap is underestimated by SCAN+$U$ and SCAN+vdW+$U$ by a much
smaller degree than PBE+$U$ and optPBE-vdW+$U$.

\begin{figure}[htpb]
  \includegraphics[width=\linewidth]{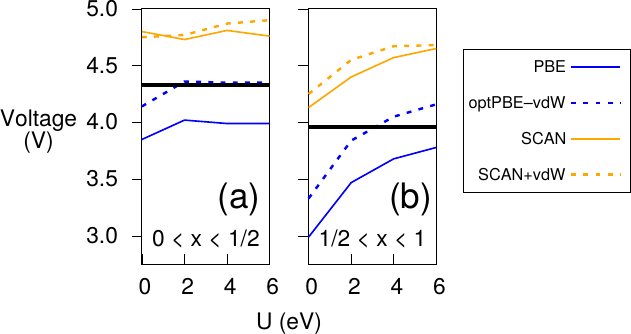}
  \caption{Li$_x$CoO$_2$ intercalation voltage for (a) $x<1/2$ and (b)
    $x>1/2$ within PBE, optPBE-vdW, SCAN, and SCAN+vdW for
    DFT+$U$. The solid black horizontal lines indicate the
    experimental voltage.\label{fig:dftu_half_voltages}}
\end{figure}

The Li$_x$CoO$_2$ voltage is significantly underestimated within pure
DFT using PBE (by 0.8 V). The significant increase in the predicted
$V$ when adding $U$ to PBE, which is dampened via a spurious
metal-insulator transition for CoO$_2$, is still insufficient to
achieve agreement with the experimental $V$
\cite{aykol_local_2014,aykol_van_2015,isaacs_compositional_2017}. As
was previously shown \cite{aykol_van_2015}, adding vdW interactions
via optPBE-vdW+$U$ further enhances $V$ with respect to PBE+$U$ and
enables agreement with experiment. Therefore, it was suggested
\cite{aykol_van_2015} that such nonlocal correlation effects were
associated with the $V$ underprediction within PBE+$U$ for
Li$_x$CoO$_2$ (and possibly other transition metal oxides). We
reproduce the previous result here and find that the optPBE-vdW+$U$
voltage agrees with experiment for $U=4.4$ eV.

SCAN provides a drastically different $V$ prediction for Li$_x$CoO$_2$
\cite{chakraborty_predicting_2018,isaacs_compositional_2019},
moderately \textit{over}estimating (by 0.3 V) the experimental
voltage. Since $U$ serves to increase $V$ in this case, the SCAN+$U$
voltage predictions for Li$_x$CoO$_2$ become even further from
experiment. As observed in many of the cases discussed above, adding
vdW interactions to SCAN+$U$ (SCAN+vdW+$U$) has a relatively modest
impact as compared to the difference between optPBE-vdW+$U$ and
PBE+$U$. SCAN+vdW+$U$ provides $V$ predictions for Li$_x$CoO$_2$ no
more than $0.1$ V larger (further from experiment) than SCAN+$U$.

Similar behavior is found in terms of the half voltages for
Li$_x$CoO$_2$, shown in Fig. \ref{fig:dftu_half_voltages}: (1) For
PBE, $U$ enhances the half voltages, but not enough to reach
experimental values, (2) optPBE-vdW+$U$ provides a substantial
increase over PBE+$U$ and enables agreement with experiment (for $U$
close to 3 eV), and (3) $U$ (vdW interactions) generally tends to
significantly (moderately) enhance the already-too-large voltages of
SCAN. We note that, despite $U$ further overestimating the voltage
magnitudes when applied to SCAN(+vdW), it does lead to an improved
voltage gap at $x=1/2$.

Although optPBE-vdW+$U$ achieves agreement with the experimental
voltage (overall the full and half $x$ ranges), it may not provide an
accurate overall description of Li$_x$CoO$_2$. As shown in the
Supplemental Material, although it exhibits an accurate prediction of
the LiCoO$_2$ band gap, optPBE-vdW+$U$ exhibits the same spurious
orderings as PBE+$U$: CoO$_2$ gap opening and large magnetic moment of
PBE+$U$, as well as Li$_{1/2}$CoO$_2$ charge ordering and gap opening.

We discuss two possible alternatives to optPBE-vdW+$U$ for best
describing Li$_x$CoO$_2$. The first alternative is to use pure SCAN.
Despite modest voltage overestimation (e.g., 0.3 V for $0 < x < 1$),
SCAN does not exhibit any of the spurious gap opening or charge
ordering discussed above. It also exhibits a very accurate LiCoO$_2$
band gap, Li$_{1/2}$CoO$_2$ magnetic moment, and reasonably accurate
volume predictions, as shown in Fig.  \ref{fig:dft_volume} and the
Supplemental Material. In addition, although the overall voltage
magnitudes are moderately overestimated, the $x=1/2$ voltage gap
(related to the $x=1/2$ formation energy) agrees decently well with
experiment, as discussed in the previous section. The second
alternative is to use DFT+DMFT (to which DFT+$U$ is a static
approximation ignoring dynamical correlations) in conjunction with
SCAN, as very recent work using non-charge-self-consistent DFT+DMFT
\cite{isaacs_compositional_2019} found that dynamical correlations (1)
are large and $x$-dependent in Li$_x$CoO$_2$, (2) help eliminate the
spurious gaps and charge ordering of DFT+$U$, and (3) reduce the
predicted $V$ such that the SCAN+DMFT voltage is likely to agree well
with experiment. Further work to assess which of these alternatives
(or another) is optimal to accurately describe Li$_x$CoO$_2$ will be
important future work.

Finally, we summarize our overall findings regarding describing
battery cathode materials within DFT+$U$. As discussed in the previous
section, within pure DFT, it is clear that (1) SCAN is superior to PBE
and (2) adding additional vdW interactions beyond those intrinsic to
SCAN is not essential and is in some cases detrimental. With DFT+$U$,
the results are less clear cut.

In the case of Li$_x$TiS$_2$, adding Hubbard $U$ generally yields no
improvement over the corresponding pure DFT $V$ results (which are
only modestly underestimated with SCAN and SCAN+vdW), if one takes
into account the spurious LiTiS$_2$ metal-insulator transition
predicted by DFT+$U$ occurring for sufficiently-large $U$. In
contrast, Hubbard $U$ is essential to achieve a voltage prediction in
agreement with experiment for Li$_x$FePO$_4$ and LiMn$_2$O$_4$.
Therefore, the new SCAN functional does not eliminate the need for
Hubbard $U$ corrections. In fact, we find SCAN+$U$ provides the best
description of these two cathode materials. Therefore, it is not true
that SCAN eliminates the need for $U$ for battery cathode materials in
general, in contrast to this conclusion reached by Chakraborty
\textit{et al.} in their study of layered systems
\cite{chakraborty_predicting_2018}. Although SCAN provides an
excellent voltage prediction for Li$_x$NiO$_2$, an improved
description can be achieved via DFT+$U$ calculations based on PBE.
Therefore, we find that calculations based on SCAN should not be
universally considered superior to those based on PBE. Finally, in the
case of Li$_x$CoO$_2$, none of the methods considered here gives a
sufficient description of both the voltage and electronic structure,
though SCAN arguably fares the best.

Taking all these results into account, despite the improved
performance obtained via pure DFT and DFT+$U$ calculations based on
SCAN for certain cases, we find that no single method can sufficiently
accurately describe the voltage and overall structural, electronic,
and magnetic properties (i.e., yielding errors no more than 5\% for
voltage, volume, band gap, and magnetic moments) of the battery
cathode materials considered here. Our results strongly motivate the
need for improved electronic structure approaches for such systems.

\section{Conclusions}\label{sec:conclusions}

Despite the great need for an accurate and computationally inexpensive
approach to characterize and design battery cathode materials, such a
method still remains out of reach at present. Within pure DFT, SCAN is
a significant improvement over PBE for describing cathode materials,
though appreciable errors remain. Methods incorporating explicit vdW
interactions are not critical and in cases even detrimental when
applied in conjunction with SCAN, which already intrinsically contains
some intermediate-range vdW interactions.

Hubbard $U$ corrections considered within DFT+$U$ are essential to
achieve an accurate voltage prediction in some cases (e.g.,
Li$_x$FePO$_4$ and Li$_x$Mn$_2$O$_4$) and detrimental in others (e.g.,
Li$_x$TiS$_2$). Although we find SCAN+$U$ provides the best
description for Li$_x$FePO$_4$ and Li$_x$Mn$_2$O$_4$, we find PBE+$U$
gives the best description for Li$_x$NiO$_2$, suggesting DFT+$U$
calculations based on SCAN should not be considered universally
superior to those based on PBE. No method here is completely
satisfactory to describe Li$_x$CoO$_2$, though the SCAN description
perhaps has the fewest deficiencies. Our results motivate the need to
develop improved electronic structure descriptions that can accurately
describe the thermodynamics and electronic structure of battery
cathode materials.

\begin{acknowledgments}
  We acknowledge support from Toyota Research Institute through the
  Accelerated Materials Design and Discovery program (development of
  software tools for automating electronic structure calculations) and
  the Center for Electrochemical Energy Science (CEES), an Energy
  Frontier Research Center funded by the U.S. Department of Energy,
  Office of the Science, Basic Energy Science under Award No.
  DE-AC02-06CH11357 (voltage calculations). Computational resources
  were provided by the National Energy Research Scientific Computing
  Center (U.S. Department of Energy Contract DE-AC02-05CH11231) and
  the Extreme Science and Engineering Discovery Environment (National
  Science Foundation Contract ACI-1548562).
\end{acknowledgments}

\bibliography{scan_battery}

\end{document}